\documentclass[12pt]{iopart}
\usepackage{iopams}

\begin{document}

\title[Energy-momentum density in small regions: the classical
pseudotensors]{Energy-momentum density in small regions: the
classical pseudotensors}

\author{Lau Loi So$^{1}$\footnote{Present address Department of Physics, Tamkang
University, Tamsui 251, Taiwan,  E-mail address:
s0242010@webmail.tku.edu.tw}, James M Nester$^{1,2,3}$ and Hsin
Chen$^{1}$}

\address{$^1$Department of Physics,
National Central University,  Chungli 320, Taiwan}
\address{$^2$Institute of Astronomy, National Central University,
Chungli 320, Taiwan}
\address{$^3$Center for Mathematics and Theoretical Physics,
National Central University, Chungli 320, Taiwan}
\ead{nester@phy.ncu.edu.tw}

%\begin{center}
%Lau Loi So\footnote{Present address Department of Physics, Tamkang
%University, Tamsui 251, Taiwan. E-mail address:
%s0242010@webmail.tku.edu.tw},\ James M. Nester\footnote{E-mail
%address: nester@phy.ncu.edu.tw}\
%and Hsin Chen\footnote{E-mail address: s2242005@cc.ncu.edu.tw}\\
%$^{1,2,3}$Department of Physics, National Central University,
%Chungli 320, Taiwan.\\

\begin{abstract}
The values for the gravitational energy-momentum density, given by
the famous classical pseudotensors: Einstein, Papapetrou,
Landau-Lifshitz, Bergmann-Thompson, Goldberg, M{\o}ller, and
Weinberg, in the small region limit are found to lowest
non-vanishing order in normal coordinates. All except M{\o}ller's
have the zeroth order material limit required by the equivalence
principle. However for small vacuum regions we find that {\it none}
of these classical holonomic pseudotensors satisfies the criterion
of being proportional to the Bel-Robinson tensor. Generalizing an
earlier work which had identified one case, we found another
independent linear combination satisfying this requirement---and
hence a one parameter set of linear combinations of the classical
pseudotensors with this desirable property.
\end{abstract}


\submitto{\CQG}

\maketitle

\section{Introduction}
The localization of energy for gravitating systems remains an
outstanding problem. Unlike all other source and interaction fields,
the standard techniques for identifying an energy-momentum density
for the gravitational field have only yielded various expressions
which are inherently reference frame dependent; these non-covariant
expressions are often referred to as pseudotensors. This
non-covariant feature can be understood as an inevitable consequence
of the equivalence principle, which precludes the detection of the
gravitational field at a point---so one cannot have a point-wise
well-defined energy-momentum density for gravitating systems (for a
good discussion of this point see Ch.~20 in \cite{MTW}).

The energy-momentum pseudotensor approach has largely been displaced
by the more modern perspective of quasi-local: energy-momentum is to
be associated with a closed 2-surface (for a review of the
quasi-local idea see \cite{Sza04}).
 One quasi-local formulation is in terms of the Hamiltonian.
 The Hamiltonian for evolving a (generally finite) spacetime region
includes a boundary term.
 Quasi-local quantities are associated with this Hamiltonian
boundary term.
 There are many possible quasi-local expressions simply because
there are many possible boundary terms.  They are all physically
meaningful, for each distinct boundary term is associated with a
distinct physical boundary condition (which is given by what must be
held fixed in the Hamiltonian variation).
 We note that this Hamiltonian quasi-local approach includes all the traditional
 pseudotensors, since
each is generated by a superpotential which serves as a special type
of Hamiltonian boundary term \cite{CN99,CNC99,CN00,beij,Nes04}. Via
the Hamiltonian formulation the ambiguities of the traditional
pseudotensors (which expression? which coordinate system?) are
clarified.  Hence, from the perspective of this Hamiltonian boundary
term approach to quasi-local energy-momentum the traditional
pseudotensors are still of interest.

For pseudotensor expressions the physical energy-momentum of the
gravitational field is inextricably bound up with the choice of
reference frame. This problem has long been recognized. Indeed soon
after Einstein had proposed his expression for gravitational energy
it was noted that with certain choices of coordinates it could give
both a nonzero energy for Minkowski space and a vanishing energy for
the Schwarzschild solution \cite{bauer,schr}. Notwithstanding this,
in many cases the reference frame ambiguity is not a
problem---simply because there is an obvious natural choice of
reference frame.

In particular in evaluating the energy-momentum for an
asymptotically flat gravitating system, the asymptotic Minkowski
space provides a natural and unambiguous reference for the
coordinate system. For this case almost all the classic
pseudotensors (except for M{\o}ller's 1958 expression \cite{Mol58})
give the standard value for the total energy-momentum. Now there is
another case where there is a natural and unambiguous reference:
namely in a small region, where one can use the flat tangent space
at some interior point to determine a Minkowski coordinate system.
Here we shall test the classical pseudotensors in this limit.

Specifically we will be concerned with the famous pseudotensors due
to Einstein \cite{Tra62}, Papapetrou \cite{Papa48,Gupta,Jackiw},
Landau-Lifshitz \cite{LL}, Bergmann-Thompson \cite{BT53},  Goldberg
\cite{Gol58},  M{\o}ller \cite{Mol58},  and Weinberg
\cite{Wein72,MTW}.
 Although in some interesting cases many of the pseudotensors
 give identical answers (see \cite{ACV96}), in the small vacuum region
limit considered here that will not be the case.

 Note that a good energy-momentum  expression for gravitating
systems should satisfy a variety of requirements (see, e.g.,
\cite{LY03,Sza04}), including giving the standard values for the
total quantities for asymptotically flat space and reducing to the
material energy momentum in the appropriate limit (equivalence
principle). No entirely satisfactory expression has yet been
identified.  One of the most restrictive requirements is positivity.

 It is generally accepted that gravitational energy should be
positive; indeed positive energy proofs have been heralded (e.g.,
\cite{SY79,Wit81,LY03}).  Positivity is difficult to prove in
general. One can regard positivity as an important test for
quasi-local energy expressions.
  One limit that is
not so difficult, and which has not been systematically investigated
for all the classical pseudotensor expressions, is the small region
limit. The small region requirements have not yet been applied to
many energy-momentum expressions. We found that they afford both
interesting restrictions and unexpected freedom.

For a small region within matter the equivalence principle requires
that the energy-momentum expression should be dominated by the
material energy-momentum tensor.  On the other hand the positivity
of gravitational energy in a small vacuum region is assured if its
Taylor series expansion in Riemann normal coordinates is at the
second order a positive multiple of the Bel-Robinson tensor.

Some time ago Deser, Franklin and Seminara \cite{DFS99} presented a
discussion of pseudotensors in a small region, which has been a
major sources of inspiration for us. In that work the main
techniques that we will use here were developed. They examined the
Taylor expansion of a pseudotensor around a preselected (vacuum)
point using Riemann normal coordinates (RNC). Considering
pseudotensors derived from superpotentials, it was  noted that the
leading order non-vanishing vacuum expansion was at second order and
would be quadratic in the Riemann tensor. They then identified four
basis expressions for such terms and related them to the famous
Bel-Robinson tensor.

They argued that it is desired that one get the Bel-Robinson tensor,
$B_{\alpha\beta\mu\nu}$. They found exactly one such expression from
a certain linear combination, $B_{\alpha\beta\mu\nu}
=\partial^2{}_{\mu\nu}(L_{\alpha\beta}+\frac{1}{2}E_{\alpha\beta})$,
of the Landau-Lifshitz and Einstein pseudotensors, a combination
which they argued was unique.

We have reexamined the issue, considering all the aforementioned
pseudotensors.   Here we report in detail on our results, which were
first obtained in \cite{So06} and briefly announced in
\cite{icga7so}. We found that  no classical holonomic pseudotensor
gives the required second order vacuum result (surely this
contributed to the difficulty in finding a positive gravitational
energy proof). However, using similar methods to the work just
mentioned, we found another independent combination of the
Bergmann-Thompson, Papapetrou and Weinberg pseudotensors, and thus a
one parameter set of pseudotensors, with the same desired
Bel-Robinson property. (This was overlooked in the earlier work
because they had not allowed for pseudotensors which require the
explicit use of the Minkowski metric as a reference).

\section{The classical pseudotensors}

The classical pseudotensors can be obtained by suitably rearranging
Einstein's equation:
\begin{equation}
G_{\mu\nu}=\kappa T_{\mu\nu},\label{einfeq}\end{equation}
(where $\kappa:=8\pi G/c^4$). One specific way is to choose a
suitable {\em superpotential} $U^{\mu\lambda}{}_\nu\equiv
U^{[\mu\lambda]}{}_\nu$ and {\em define} the associated
gravitational energy-momentum density pseudotensor by
\begin{equation}
2\kappa t^\mu{}_\nu:=\partial_\lambda
U^{\mu\lambda}{}_\nu-2|g|^{\frac12}G^\mu{}_\nu. \label{paradigm}
\end{equation}
Einstein's equation now takes the form
\begin{equation}
2\kappa {\cal T}^\mu{}_\nu:=2\kappa(|g|^{\frac12}T^\mu{}_\nu
+t^\mu{}_\nu)=\partial_\lambda U^{\lambda\mu}{}_\nu.
\end{equation}
Because of the antisymmetry of the superpotential the total
energy-momentum density complex is automatically conserved:
$\partial_\mu {\cal T}^\mu{}_\nu\equiv0$.

 There are some variations on the idea, the classical pseudotensorial total energy-momentum density
complexes all follow from associated superpotentials according to
one of the patterns
\begin{equation}
2\kappa {\cal T}^\mu{}_\nu=\partial_\lambda
U^{\mu\lambda}{}_\nu,\quad
2\kappa {\cal T}^{\mu\nu}=\partial_\lambda U^{\mu\lambda\nu},%
\quad 2\kappa {\cal
T}^{\mu\nu}=\partial_{\alpha\beta}H^{\alpha\mu\beta\nu},
\end{equation}
where the superpotentials have certain symmetries which
automatically guarantee conservation: specifically
$U^{\mu\lambda}{}_\nu\equiv U^{[\mu\lambda]}{}_\nu$,
 $U^{\mu\lambda\nu}\equiv
U^{[\mu\lambda]\nu}$, while $H^{\alpha\mu\beta\nu}$ has the
symmetries of the Riemann tensor. In particular the Einstein total
energy-momentum density follows from the Freud superpotential
\cite{Freud}
\begin{equation}
U_{\rm F}^{\mu\lambda}{}_\nu:=-|g|^{\frac12}g^{\beta\sigma}
\Gamma^\alpha{}_{\beta\gamma}\delta^{\mu\lambda\gamma}_{\alpha\sigma\nu};\label{UF}
\end{equation}
 while the Bergmann-Thompson \cite{BT53}, Landau-Lifshitz \cite{LL}, Papapetrou \cite{Papa48}, Weinberg \cite{Wein72} and
 M{\o}ller \cite{Mol58}
 expressions can be obtained from the respective superpotentials
\begin{eqnarray}
U_{\rm BT}^{\mu\lambda\nu}&:=&g^{\nu\delta}U_{\rm
F}^{\mu\lambda}{}_\delta,\label{UBT}\\
U_{\rm LL}^{\mu\lambda\nu}&:=&|g|^{\frac12}U_{\rm
BT}^{\mu\lambda\nu},\quad {\mathrm{equivalently}} \quad H_{\rm
LL}^{\alpha\mu\beta\nu}:=|g|\delta^{\mu\alpha}_{ma}
g{}^{a\beta}g^{m\nu},\label{ULL}\label{HLL}\\
H_{\rm
P}^{\alpha\mu\beta\nu}&:=&\delta^{\mu\alpha}_{ma}\delta^{\nu\beta}_{nb}\bar
g{}^{ab}(|g|^{\frac12}g^{mn}),\label{HP}\\
H_{\rm
W}^{\alpha\mu\beta\nu}&:=&\delta^{\mu\alpha}_{ma}\delta^{\nu\beta}_{nb}|\bar
g|^{\frac12}\bar g{}^{ab}(-{\bar g}^{mc}{\bar g}^{nd}+{\frac12}{\bar
g}^{mn}{\bar g}^{cd}) g_{cd},\label{HW}\\
U_{\rm
M}^{\mu\lambda}{}_\nu&:=&-|g|^{\frac12}g^{\beta\sigma}\Gamma^\alpha{}_{\beta\nu}\delta^{\mu\lambda}_{\alpha\sigma}
\equiv |g|^{\frac12}g^{\beta\mu}g^{\lambda\delta}(\partial_\beta
g_{\delta\nu}-\partial_\delta g_{\beta\nu}) \label{UM}.
\end{eqnarray}
(Note that all indicies in this work refer to spacetime and range
from 0 to 3, otherwise  our conventions follow \cite{MTW}.)

These expressions are all non-covariant.  As they depend on the
coordinates in a non-tensorial way, they can at best be expected to
give sensible energy-momentum values only in certain
coordinates---which are in some suitable sense nearly Minkowski
coordinates. Given that we have such coordinates we also have an
underlying Minkowski space reference structure.  Some of the
superpotentials explicitly include this reference metric, which here
 has the Minkowski values $\bar g_{ij}={\rm diag}(-1,+1,+1,+1)$.
Our basic philosophy is that energy-momentum is properly a covector,
and hence properly the energy-momentum density should be a weight
one density with the index positions $t^\mu{}_\nu$.  For all of the
classical pseudotensors  this can be achieved by introducing
suitable factors of the Minkowski metric and its determinant.
According to this philosophy only the Einstein and M{\o}ller
expressions have proper expressions that do not explicitly need the
Minkowski metric associated with the chosen coordinates.

It should be noted that, while in general identifying physically
meaningful Minkowski coordinates is problematical, in the small
region case of interest to us here at any chosen point there is a
natural local Minkowski structure, as we discuss in the next
section.

\section{Some Technical background}

\subsection{Riemann normal coordinates (RNC)}

As Riemann first argued (see, e.g., Spivak \cite{Spivak}), at any
preselected point one can choose coordinates such that at the point
$x^\mu=0$, the metric coefficients have the standard flat values,
the first derivatives of the metric vanish, and the second
derivatives have the minimum number (20 for $n=4$) of independent
values. Specifically
\begin{equation}\fl\qquad
g_{\alpha\beta}|_0=\bar g_{\alpha\beta}, \quad \partial_\mu
g_{\alpha\beta}|_0=0, \quad
3\partial_{\mu\nu}g_{\alpha\beta}|_0=-(R_{\alpha\mu\beta\nu}+R_{\alpha\nu\beta\mu})|_0,
\end{equation}
where $R^\alpha{}_{\beta\mu\nu}$ is the Riemannian curvature tensor%
%(all indicies are holonomic, our conventions, unless otherwise specified, follow MTW \cite{MTW})
, and, in our case, $\bar
g_{\alpha\beta}=\eta_{\alpha\beta}=\hbox{diag}(-1,+1,+1,+1)$ is the
Minkowski spacetime metric.  The corresponding Levi-Civita
connection values are
\begin{equation}
\Gamma^\alpha{}_{\beta\gamma}|_0=0,\qquad 3\partial_\mu
\Gamma^\alpha{}_{\beta\nu}|_0=-(R^\alpha{}_{\beta\nu\mu}+R^\alpha{}_{\nu\beta\mu})|_0.
\end{equation}

\subsection{Quadratic curvature basis}
It turns out that when expanded in RNC the lowest non-vanishing
vacuum energy-momentum expressions are of the second order and are
quadratic in the curvature tensor:
$t_{\mu\nu}\sim(R_{\cdot\cdot\cdot\cdot}R^\cdot{}_\cdot{}^\cdot{}_\cdot)_{\mu\nu
ij}x^ix^j$. An investigation \cite{DFS99} of all such possible terms
(taking into account the Weyl = vacuum Riemann tensor symmetries)
shows that they can be written in terms of
\begin{equation}
Q_{\mu\alpha\nu\beta}:=R_{a\mu b\alpha}R^a{}_\nu{}^b{}_\beta\equiv
Q_{\nu\beta\mu\alpha}\equiv Q_{\alpha\mu\beta\nu},
\end{equation}
where all the symmetries have been indicated. The cited work defined
the three basis combinations
\begin{equation}\fl\qquad
X_{\mu\nu\alpha\beta}:=2Q_{\alpha(\mu\nu)\beta}, \quad
Y_{\mu\nu\alpha\beta}:=2Q_{\alpha\beta(\mu\nu)}, \quad
Z_{\mu\nu\alpha\beta}:=Q_{\alpha\mu\beta\nu} +
Q_{\alpha\nu\beta\mu},
\end{equation}
along with the trace tensor
\begin{equation}
T_{\mu\nu\alpha\beta}=-\frac16
g_{\mu\nu}Q^\sigma{}_{\alpha\beta\sigma}.
\end{equation}
One could write all our vacuum expansions to second order as linear
combinations of $X$, $Y$, $Z$, $T$.  However it is more suitable for
physical purposes to use the Bel-Robinson tensor.

\subsection{The Bel-Robinson and two other tensors}
The Bel-Robinson tensor
\begin{equation}
B_{\mu\nu\alpha\beta}=R_{\rho\mu\sigma\alpha}R^\rho{}_\nu{}^\sigma{}_\beta+R_{\rho\mu\sigma\alpha}R^\rho{}_\nu{}^\sigma{}_\beta
-\frac12g_{\mu\nu}R_{\alpha\rho\sigma\tau}R_\beta{}^{\rho\sigma\tau}
\end{equation}
has many well known remarkable properties, see e.g.,
\cite{DFS99,Gar01}.  For our considerations we are interested in it
only in the vacuum, where the Riemann tensor reduces to the Weyl
tensor. In this case the Bel-Robinson tensor is completely symmetric
and traceless, and the last term admits an alternate form using
\begin{equation}4R_{\alpha\rho\sigma\tau}R_\beta{}^{\rho\sigma\tau}
=g_{\alpha\beta}R_{\kappa\lambda\gamma\delta}R^{\kappa\lambda\gamma\delta}.\label{traceid}\end{equation}
In vacuum the Bel-Robinson tensor $B$ and two other convenient
tensors $S$ and $K$
 are given by
\begin{eqnarray}
B_{\alpha\beta\mu\nu}
&:=&R_{\alpha\lambda\mu\sigma}R_{\beta}{}^\lambda{}_{\nu}{}^{\sigma}
+R_{\alpha\lambda\nu\sigma}R_{\beta}{}^{\lambda}{}_{\mu}{}^{\sigma}
-\frac{1}{8}g_{\alpha\beta}g_{\mu\nu}R_{\kappa\lambda\gamma\delta}R^{\kappa\lambda\gamma\delta},  \\
S_{\alpha\beta\mu\nu}
&:=&R_{\alpha\mu\lambda\sigma}R_{\beta\nu}{}{}^{\lambda\sigma}
+R_{\alpha\nu\lambda\sigma}R_{\beta\mu}{}{}^{\lambda\sigma} +\frac14
g_{\alpha\beta}g_{\mu\nu}R_{\kappa\lambda\gamma\delta}R^{\kappa\lambda\gamma\delta},\\
K_{\alpha\beta\mu\nu}&:=&R_{\alpha\lambda\beta\sigma}R_\mu{}^\lambda{}_\nu{}^\sigma+R_{\alpha\lambda\beta\sigma}R_\nu{}^\lambda{}_\mu{}^\sigma
-\frac38g_{\alpha\beta}g_{\mu\nu}R_{\kappa\lambda\gamma\delta}R^{\kappa\lambda\gamma\delta}.
\end{eqnarray}
In term of the aforementioned basis for quadratic terms
\begin{equation}
B=Z+3T, \qquad S=-2X+2Z-6T, \qquad K=Y+9T.
\end{equation}
(Here and below we have suppressed some obvious indicies.)

Our leading order non-vanishing vacuum expressions will be given as
linear combinations of $B$, $S$, and $K$. As can be directly
verified, these three combinations (by virtue of (\ref{traceid}))
satisfy the divergence free condition,
\begin{equation}
\partial_\beta (x^i x^j t_{ij}{}^{\alpha\beta})\equiv2 x^j t_{ij}{}^{\alpha i}\equiv0,
\end{equation}
and all such tensors are some linear combination of these three.
While the tensor $S$ has been known for a long time, we here draw
attention to the tensor $K$ which also enjoys this divergence free
property.

%%%%%%%%%%%%%%%%%%%%%%%%%%%%%%%%%%%%%%%%%%%%%%%%%%%%%%%%%%%%%%%%%%%%
\section{The small region limit}

Here the small region values for the classical pseudotensors are
presented.

\subsection{Einstein}

The Einstein total energy-momentum complex can be obtained from the
Freud superpotential (\ref{UF}):
\begin{eqnarray}\fl\qquad\quad
2\kappa {\cal T}_{\rm E}^\mu{}_\nu&:=&\partial_\lambda U_{\rm
F}^{\mu\lambda}{}_\nu:=\partial_\lambda(-|g|^{\frac12}g^{\beta\sigma}
\Gamma^\alpha{}_{\beta\gamma}\delta^{\mu\lambda\gamma}_{\alpha\sigma\nu})\qquad
\\
\fl&\equiv&-|g|^{\frac12}g^{\beta\sigma}
({\frac12}R^\alpha{}_{\beta\lambda\gamma}-\Gamma^\alpha{}_{\delta\lambda}\Gamma^\delta{}_{\beta\gamma})\delta^{\mu\lambda\gamma}_{\alpha\sigma\nu}
-
\partial_\lambda(|g|^{\frac12}g^{\beta\sigma})
\Gamma^\alpha{}_{\beta\gamma}\delta^{\mu\lambda\gamma}_{\alpha\sigma\nu}\qquad\\
&\equiv&2|g|^{\frac12} G^\mu{}_\nu-
|g|^{\frac12}[\Gamma^\delta{}_{\delta\lambda}\Gamma^{\alpha\sigma}{}_\gamma-\Gamma^{\sigma\beta}{}_\lambda\Gamma^\alpha{}_{\beta\gamma}
]\delta^{\mu\lambda\gamma}_{\alpha\sigma\nu}.\qquad
\end{eqnarray}
Using the Einstein field equation (\ref{einfeq}) this relation takes
the form
\begin{equation}
{\cal T}_{\rm E}^\mu{}_\nu=|g|^{\frac12}T^\mu{}_\nu+t_{\rm
E}^\mu{}_\nu,\label{eincmplx}
\end{equation}
where the Einstein energy-momentum pseudotensor density is
\begin{equation}t_{\rm E}^\mu{}_\nu:=-(2\kappa)^{-1}
|g|^{\frac12}[\Gamma^\delta{}_{\delta\lambda}\Gamma^{\alpha\sigma}{}_\gamma-\Gamma^{\sigma\beta}{}_\lambda\Gamma^\alpha{}_{\beta\gamma}
]\delta^{\mu\lambda\gamma}_{\alpha\sigma\nu}.
\end{equation}
 Expanding in
normal coordinates, to zeroth order the pseudotensor vanishes, so
(\ref{eincmplx}) reduces to ${\cal T}_{\rm E}^\mu{}_\nu=
|g|^{\frac12}T^\mu{}_\nu$, which is the matter interior limit
expected from the equivalence principle. In vacuum ${\cal T}_{\rm
E}=t_{\rm E}$, and the first non-vanishing contribution appears in
the second order (note: all trace $\Gamma$ terms are proportional to
Ricci and hence vanish in vacuum); after some computation we find in
vacuum
\begin{equation}\fl\quad\quad
2\kappa {\cal T}_{\rm E}^\mu{}_\nu=2\kappa t_{\rm E}^\mu{}_\nu
=|g|^{\frac12}(-2\Gamma^{\mu\beta}{}_\alpha\Gamma^\alpha{}_{\beta\nu}+\delta^\mu_\nu\Gamma^{\sigma\beta}{}_\alpha\Gamma^\alpha{}_{\beta\sigma}
) \simeq{\frac{x^ix^j}{2\cdot3\cdot3}}(4B-S)_{ij}{}^\mu{}_\nu,\quad
\end{equation}
a result known for some time \cite{MTW,Gar73,DFS99,Gar01}.

\subsection{Bergmann-Thompson, Landau-Lifshitz and Goldberg}

The total energy-momentum complex proposed by Bergmann and Thompson
\cite{BT53} can be obtained from a transvected version of the Freud
superpotential (\ref{UBT}):
\begin{equation}
2\kappa {\cal T}_{\rm BT}^{\mu\nu}:=\partial_\lambda(
g^{\nu\delta}U_{\rm F}^{\mu\lambda}{}_\delta) \equiv 2\kappa {\cal
T}_{\rm E}^\mu{}_\delta g^{\nu\delta} -|g|^{\frac12}g^{\beta\sigma}
\Gamma^\alpha{}_{\beta\gamma}\delta^{\mu\lambda\gamma}_{\alpha\sigma\delta}\partial_\lambda
g^{\nu\delta}.\label{BTcmplx}
\end{equation}
Consequently the Bergmann-Thompson gravitational energy-momentum
pseudotensor density, found from a decomposition of the form ${\cal
T}_{\rm BT}=|g|^{\frac12}T+t_{\rm BT}$, is
\begin{equation}
t_{\rm BT}^{\mu\nu}=t_{\rm E}^\mu{}_\delta g^{\nu\delta}
-|g|^{\frac12}g^{\beta\sigma}
\Gamma^\alpha{}_{\beta\gamma}\delta^{\mu\lambda\gamma}_{\alpha\sigma\delta}\partial_\lambda
g^{\nu\delta}.
\end{equation}
In normal coordinates to zeroth order the pseudotensor $t_{\rm BT}$
vanishes so ${\cal T}_{\rm BT}=|g|^{\frac12}T$, which is the desired
limit inside of matter. In vacuum the complex (\ref{BTcmplx})
reduces to the pseudotensor, which has its first nonvanishing
contribution at the second order:
\begin{equation}\fl\quad
2\kappa t_{\rm
BT}^{\mu\nu}=|g|^{\frac12}\left[2\Gamma^{\mu}{}_{\lambda\sigma}
\Gamma^{[\nu\lambda]\sigma} -2\Gamma_{\lambda\sigma}{}^{\mu}
\Gamma^{(\nu\sigma)\lambda}
+g^{\mu\nu}\Gamma_{\lambda\sigma\tau}\Gamma^{\sigma\lambda\tau}\right]
\simeq
{\frac{x^ix^j}{2\cdot3\cdot3}}(7B+{\frac12}S)_{ij}{}^{\mu\nu}.\
\label{BTvac}
\end{equation}

The better known Landau-Lifshitz expression \cite{LL} can be
obtained from this transvected superpotential with an additional
improper density weight factor (\ref{ULL}):
\begin{equation}\fl\qquad
2\kappa {\cal T}_{\rm LL}^{\mu\nu}:=\partial_\lambda({|g|}^{\frac12}
g^{\nu\delta}U_{\rm F}^{\mu\lambda}{}_\delta) \equiv 2\kappa
|g|^{\frac12}{\cal T}_{\rm E}^{\mu\nu}-|g|^{\frac12}g^{\beta\sigma}
\Gamma^\alpha{}_{\beta\gamma}\delta^{\mu\lambda\gamma}_{\alpha\sigma\delta}\partial_\lambda
(|g|^{\frac12}g^{\nu\delta}).
\end{equation}
A short calculation shows that this extra weight factor makes no
contribution to the zeroth-order matter-interior limit nor to the
second order result in vacuum. Thus the Landau-Lifshitz vacuum
result is  given by (\ref{BTvac}), as was noted some time ago
\cite{DFS99}.

By the way, Goldberg \cite{Gol58} has proposed an infinite class of
arbitrary density weight energy-momentum expressions obtained from
those of Einstein and Landau-Lifshitz in a manner similar to that
just considered. For essentially the same reasons, Goldberg's
weighted density factors will not yield any modified values in the
small region limits we are considering here.

\subsection{Papapetrou}

For our analysis of the Papapetrou \cite{Papa48} and Weinberg
\cite{Wein72} expressions we use the Einstein tensor expansion
\begin{eqnarray}\fl\quad
2G^{\mu\nu}&\equiv&\delta^{\mu\alpha}_{ma}\delta^{\nu\beta}_{nb}g^{ab}
\left[{\textstyle\frac12}g^{ef}g^{mn}-g^{me}g^{nf}\right]\partial_{\alpha\beta}g_{ef}\quad\quad\quad\nonumber\\
\fl\quad&+&
(2g^{\beta\mu}g^{\gamma\nu}-g^{\mu\nu}g^{\beta\gamma})\left(\partial_\alpha
g^{\alpha\delta}\Gamma_{\delta\beta\gamma}-\partial_\gamma
g^{\alpha\delta}\Gamma_{\delta\beta\alpha}+\Gamma^\alpha{}_{\delta\alpha}\Gamma^\delta{}_{\beta\gamma}
-\Gamma^\alpha{}_{\delta\gamma}\Gamma^\delta{}_{\beta\alpha}\right).\quad\quad
\label{Einsteintensor}
\end{eqnarray}
For the Papapetrou energy-momentum complex
\cite{Papa48,Gupta,Jackiw} we find, using (\ref{Einsteintensor}) and
the appropriate superpotential (\ref{HP}),
\begin{eqnarray}\fl\quad
2\kappa {\cal T}^{\mu\nu}_{\rm
P}&:=&\delta^{\mu\alpha}_{ma}\delta^{\nu\beta}_{nb}\bar
g{}^{ab}\partial_{\alpha\beta}(|g|^{\frac12}g^{mn})\\
\fl\quad&\equiv&\delta^{\mu\alpha}_{ma}\delta^{\nu\beta}_{nb}
(g{}^{ab}-\Delta g^{ab})\partial_\alpha
\Bigl[|g|^{\frac12}({\frac12}g^{ef}g^{mn}-g^{me}g^{nf})\partial_\beta g_{ef}\Bigr]\\
\fl\quad&\simeq&2|g|^{\frac12}G^{\mu\nu}+|g|^{\frac12}
(2g^{\beta\mu}g^{\gamma\nu}-g^{\mu\nu}g^{\beta\gamma})\left[\partial_\gamma
g^{\alpha\delta}\Gamma_{\delta\beta\alpha}
+\Gamma^\alpha{}_{\delta\gamma}\Gamma^\delta{}_{\beta\alpha}\right]\nonumber\\
\fl\quad&&+\delta^{\mu\alpha}_{ma}\delta^{\nu\beta}_{nb} \Delta
g^{ab}|g|^{\frac12}g^{me}g^{nf}
\partial_{\alpha\beta}
g_{ef}\nonumber\\
\fl\quad&&+\delta^{\mu\alpha}_{ma}\delta^{\nu\beta}_{nb}
g{}^{ab}|g|^{\frac12}\Bigl(-{\frac12}g^{ei}g^{fj}g^{mn}
+g^{mi}g^{ej}g^{nf}+g^{me}g^{ni}g^{fj}\Bigr)\partial_\alpha
g_{ij}\partial_\beta g_{ef},\quad\quad\label{papaexp}
\end{eqnarray}
where we use for any  metric expression $\Delta F:=F-\bar F$. Here
we have indicated one way to arrange things so that we can get the
terms of the desired orders. In normal coordinates to zeroth order
within matter the expansion (\ref{papaexp}) clearly reduces to the
desired result. In vacuum to second order a lengthy calculation
yields
\begin{equation}
2\kappa t^{\mu\nu}_{\rm P}\simeq
{\frac{x^ix^j}{3\cdot3}}[4B-S-K]_{ij}{}^{\mu\nu}.
\end{equation}

\subsection{Weinberg}
With the help of (\ref{Einsteintensor}) and the appropriate superpotential (\ref{HW}) for the Weinberg \cite{Wein72}
expression (note: the same expression is considered in \cite{MTW} \S
20.2) we find
\begin{eqnarray}\fl\quad
&&2\kappa {\cal T}^{\mu\nu}_{\rm W}
:=\delta^{\mu\alpha}_{ma}\delta^{\nu\beta}_{nb}|\bar
g|^{\frac12}\bar g{}^{ab}(-{\bar g}^{me}{\bar g}^{nf}+{\frac12}{\bar
g}^{mn}{\bar g}^{ef})
\partial_{\alpha\beta}g_{ef}\\
\fl\quad&&\quad\equiv\delta^{\mu\alpha}_{ma}\delta^{\nu\beta}_{nb}|g|^{\frac12}
g{}^{ab}( -g^{me} g^{nf}+{\frac12} g^{mn} g^{ef})\partial_{\alpha\beta}g_{ef}\nonumber\\
\fl\quad&&\quad-\delta^{\mu\alpha}_{ma}\delta^{\nu\beta}_{nb}\Delta\left[|g|^{\frac12}
g^{ab}(-g^{me} g^{nf}+{\frac12} g^{mn} g^{ef})\right]\partial_{\alpha\beta}g_{ef}\\
\fl\quad&&\quad\simeq 2|g|^{\frac12}G^{\mu\nu}+|g|^{\frac12}
(2g^{\beta\mu}g^{\gamma\nu}-g^{\mu\nu}g^{\beta\gamma})\left[\partial_\gamma
g^{\alpha\delta}\Gamma_{\delta\beta\alpha}
+\Gamma^\alpha{}_{\delta\gamma}\Gamma^\delta{}_{\beta\alpha}\right]\\
\fl\quad
&&-\delta^{\mu\alpha}_{ma}\delta^{\nu\beta}_{nb}|g|^{\frac12}
\Bigl[g^{ai}g^{bj}
 g^{me} g^{nf}+g^{ab} (g^{mi}g^{ej} g^{nf}+g^{me}  g^{ni}g^{fj}
-{\frac12} g^{mn} g^{ei}g^{fj})\Bigr]\Delta
g_{ij}\partial_{\alpha\beta}g_{ef}.\nonumber
\end{eqnarray}
Here we have indicated one way to get the desired orders. In normal
coordinates, as expected, to zeroth order it gives the desired
material limit; in vacuum to second order after a long calculation
we found
\begin{equation}
2\kappa t^{\mu\nu}_{\rm W}\simeq {\frac{x^ix^j}
{3\cdot3}}[-B-2S-3K]_{ij}{}^{\mu\nu}.
\end{equation}

\subsection{M{\o}ller}

M{\o}ller's holonomic energy-momentum complex \cite{Mol58} follows
from the superpotential (\ref{UM}).  We find
\begin{eqnarray}\fl\quad
2\kappa {\cal T}_{\rm M}^\mu{}_\nu
&:=&-\partial_\lambda(|g|^{\frac12}g^{\beta\sigma}\Gamma^\alpha{}_{\beta\nu})\delta^{\mu\lambda}_{\alpha\sigma}
\equiv\partial_\lambda\left[|g|^{\frac12}g^{\beta\mu}g^{\lambda\delta}(\partial_\beta
g_{\delta\nu}-\partial_\delta g_{\beta\nu})\right]\\
\fl\quad&\equiv&|g|^{\frac12}g^{\beta\mu}g^{\lambda\delta}(\partial_{\lambda\beta}
g_{\delta\nu}-\partial_{\lambda\delta} g_{\beta\nu})
+\partial_\lambda\left[|g|^{\frac12}g^{\beta\mu}g^{\lambda\delta}\right](\partial_\beta
g_{\delta\nu}-\partial_\delta g_{\beta\nu})\\
\fl\quad&\simeq&|g|^{\frac12}R^\mu{}_\nu-\frac12g^{\mu\beta}g^{\delta\lambda}
\left(\partial_{\lambda\delta}g_{\beta\nu}-\partial_{\lambda\beta}g_{\delta\nu}+\partial_{\delta\nu}g_{\lambda\beta}
-\partial_{\beta\nu}g_{\lambda\delta}\right)+O(\Gamma^2) \\
\fl\quad &\simeq&|g|^{\frac12}R^\mu{}_\nu+O(x^2).
\end{eqnarray}
Note that the M{\o}ller energy-momentum complex fails to give the
correct small region material limit (by the way, it also fails to
agree with the proper asymptotic limit presented in \cite{MTW}, \S
20.2). Using Einstein's equation and (\ref{paradigm}) we have to
zeroth order
\begin{equation}
2t_{\rm M}^\mu{}_\nu=-(T^\mu{}_\nu+\frac12 \delta^\mu_\nu
T^\delta{}_\delta),
\end{equation}
which is non-vanishing inside of matter. Therefore the M{\o}ller
expression does not satisfy the {\it equivalence principle}. From
our perspective M{\o}ller's holonomic expression is thus {\em
disqualified} as a satisfactory description of energy-momentum.

Although it is consequently only of academic interest, nevertheless,
for completeness, we briefly report here on the small region vacuum
limit of M{\o}ller's pseudotensor. From a long calculation, which
requires the 4th order Riemann normal coordinate expansion for the
metric \cite{Ni},
\begin{equation}
g_{\alpha\beta,\sigma\lambda\mu\nu}={\rm
P}(-\frac{1}{20}R_{\alpha\sigma\beta\lambda;\mu\nu}
+\frac{2}{45}R^{\xi}{}_{\sigma\beta\lambda}R_{\xi\mu\alpha\nu}),
\end{equation}
where $\rm P$ is the $\sigma\lambda\mu\nu$ symmetrization projection
operator, we found to second order
\begin{equation}
2\kappa t_{\rm M}^\mu{}_\nu\simeq
\frac{x^ix^j}{3\cdot3}[2B-\frac{1}{2}S-K]_{ij}{}^\mu{}_\nu.
\end{equation}

\section{Combinations of the classical pseudotensors}
The M{\o}ller expression is ruled out of our further considerations
by its material and asymptotic limit. The other four expressions are
satisfactory asymptotically and in the material limit.  However none
of them give a good small vacuum value. Following the example of
\cite{DFS99} we shall consider linear combinations of the four
satisfactory classical pseudotensors.  In the aforementioned work it
was noted that in vacuum a certain combination of the Einstein and
the Landau-Lifshitz (or equivalently the Bergmann-Thompson)
expressions is to second order proportional to the Bel-Robinson
tensor; specifically they found
\begin{equation}
B_{\alpha\beta\mu\nu}=\partial^2_{\mu\nu}(t_{\rm
E}+\textstyle{\frac12}t_{\rm LL})_{\alpha\beta}.
\end{equation}

This raises a couple of issues not addressed in that earlier work.
In particular one should pay attention to the overall normalization
in order to satisfy the small region matter interior limit (the same
normalization will also give the correct magnitude for the total
energy momentum of an asymptotically flat space).  One also must
cope with the mixed index positions which naturally occur for the
Einstein pseudotensor in contrast with the two contravariant indices
of the other three pseudotensors. Algebraically at the RNC origin it
matters not whether one raises the index on the Einstein expression
with $g^{\alpha\beta}$ or ${\bar g}{}^{\alpha\beta}$. However only
the latter choice is allowed if we wish to preserve the conservation
property.  Thus we see another reason why we are paralyzed unless we
allow for the explicit appearance of the Minkowski metric in RNC.
Hence we {\it define}
\begin{equation}
t^{\mu\nu}_{\rm E}:={\bar g}{}^{\nu\gamma}t^\mu_{\rm E}{}_\gamma.
\end{equation}

With this convention we can consider the linear combinations
(suppressing indicies for simplicity)
\begin{eqnarray}\fl\quad
2\kappa t&:=&2\kappa[et_{\rm E}+bt_{\rm BT}+pt_{\rm P}+wt_{\rm
W}]\simeq(e+b+p+w)2G\\
\fl\quad&&\!\!\!
+{\frac19}xx\left[B(2e+{\frac72}b+4p-w)-S({\frac12}e-{\frac14}b+p+2w)-K(p+3w)\right],\nonumber
\end{eqnarray}
where we use $\simeq$ to mean that only the zeroth order and second
order vacuum values have been indicated. In order to have the
correct asymptotic and material limits one must require $e+b+p+w=1$,
and for a good small vacuum limit to Bel-Robinson the coefficients
of $S$ and $K$ should vanish---thus three restrictions on the four
parameters. A convenient way to parameterize the set of acceptable
coefficients is
\begin{equation}
e=1-\lambda, \quad b=2/3,\quad p=3\lambda/2-1, \quad
w=-\lambda/2+1/3,
\end{equation}
which yields
\begin{eqnarray}
 t(\lambda)&=&(1-\lambda)t_{\rm E}+\frac23t_{\rm BT}+\left(\frac{\lambda}{2}-\frac13\right)(3t_{\rm
P}-t_{\rm W})\\
&=&\left[t_{\rm E}+\frac23t_{\rm BT}-\frac13(3t_{\rm P}-t_{\rm
W})\right] +\lambda\left[ -t_{\rm E}+\frac12(3t_{\rm
P}-t_{\rm W})\right]\label{2terms}\\
&\simeq&(2\kappa)^{-1}\left[2G+\lambda Bxx/2\right]=T+\lambda
Bxx/4\kappa.
\end{eqnarray}
The combination in the first bracket of (\ref{2terms}) gives both
the desired asymptotic and small region material result, but
vanishes to 2nd order in the small vacuum region limit; the
combination in the second bracket vanishes both asymptotically and
in the material limit, while in the small vacuum limit it gives the
desired 2nd order Bel-Robinson contribution. {\it Formally} we can
choose any value for $\lambda$. Physically we want $\lambda>0$. We
know of no principle to fix the magnitude of $\lambda$.  Some values
stand out: $\lambda=2/3$ gives the (here properly normalized) case
found earlier \cite{DFS99}:
\begin{equation}
t(2/3)=(1/3)t_{\rm E}+(2/3) t_{\rm BT}\simeq T+Bxx/6\kappa.
\end{equation}
The choice $\lambda=1$ gives another simple case:
\begin{equation}
t(1)=(2/3)t_{\rm BT}+(1/6)(3t_{\rm P}-t_{\rm W})\simeq
T+Bxx/4\kappa.
\end{equation}

Note that the earlier work found just one expression.  The
restricted form of the expressions considered there (i.e., not
explicitly containing $\bar g$) are not in our opinion justified. To
take linear combinations we need to get the indicies on the Einstein
pseudotensor at the same level, that requires $\bar g_{\mu\nu}$.
Moreover, the Landau-Lifshitz expression is actually of the wrong
density weight (this can be adjusted by including a suitable power
of $|\bar g|$). To the order considered in that work one can equally
well use the Bergmann-Thompson expression, which we have used here.

Comparing with the earlier work, we note that both the Weinberg and
Papapetrou expressions cannot be constructed without explicitly
using a reference.  We certainly want to allow for these
expressions---especially the latter which has been generalized to
define the total energy for an asymptotically anti-de Sitter space
\cite{AD82}.

\section{Conclusion}

In this work we found the values for the gravitational
energy-momentum density given by the famous classical pseudotensors:
Einstein, Papapetrou, Landau-Lifshitz, Bergmann-Thompson, Goldberg,
M{\o}ller, and Weinberg, in the small region limit to lowest
non-vanishing order in normal coordinates. All except M{\o}ller's
were found to have the zeroth order material limit required by the
equivalence principle. For small vacuum regions we found that {\it
none} of these classical holonomic pseudotensors satisfies the 2nd
order vacuum criterion of being proportional to the Bel-Robinson
tensor.

However certain linear combinations of these classical holonomic
pseudotensors do give the desired 2nd order vacuum result.
Generalizing an earlier work \cite{DFS99} which had identified one
case, we found another independent linear combination satisfying
this requirement---and hence a one parameter set of linear
combinations of the Einstein, Bergmann-Thompson, Papapetrou and
Weinberg pseudotensors with this same desired property.  (This had
been overlooked because the earlier work had not allowed for the
explicit use of the Minkowski metric as a reference in the
pseudotensor.)

Regarding the physical meaning of these combinations, although that
has not been worked out in detail, there is a straightforward way to
find it. Each viable parameter choice fixes a specific Hamiltonian
boundary term---and thereby not just the value of the Hamiltonian
but also an associated boundary condition.  Nevertheless these
combinations may still appear rather artificial, as being
mathematically and physically contrived. This situation may be
contrasted with the case of another (nonholonomic) energy-momentum
expression: elsewhere we have shown that the tetrad-teleparallel
energy-momentum gauge current expression \cite{AGP00} {\em
naturally} has the desired Bel-Robinson property \cite{SN08}.

Although it still allows for a certain amount of freedom, the small
vacuum region Bel-Robinson positivity requirement provides a strong
restriction which excludes many otherwise satisfactory
energy-momentum expressions.

\ack We would like to thank the Taiwan NSC for their financial
support under the grant numbers NSC 93-2112-M-008-001,
94-2112-M008-038, 95-2119-M008-027, 96-2112-M-008-005,
97-2112-M-008-001.   JMN was also supported in part by the National
Center of Theoretical Sciences.

\end{document}